\documentclass[preprint,aps,jap,showpacs,altaffilletter,groupedaddress,frontmatterverbose]{revtex4}
\usepackage{amsmath}
\usepackage{amssymb}
\usepackage{graphicx}
\usepackage{dcolumn}
\usepackage{bm}
\usepackage{float}

\DeclareGraphicsExtensions{.pdf,.png,.jpg}
\begin{document}

\preprint{AIP/123-QED}
\title{ Correlation between magnetic interactions and domain structure in A1
FePt ferromagnetic thin films}
\author{N. \'{A}lvarez}
\author{E. Sallica Leva}
\author{R. C. Valente}
\author{M. V\'{a}squez Mansilla}
\author{J. G\'{o}mez}
\affiliation{Centro At\'{o}mico Bariloche (CNEA) and Conicet, 8400 Bariloche, R\'{\i}o
Negro, Argentina.}
\author{J. Milano}
\author{A. Butera}
\email{butera@cab.cnea.gov.ar}
\altaffiliation[Also at ]{INN - Instituto de Nanociencia y Nanotecnolog\'{\i}a }
\affiliation{Centro At\'{o}mico Bariloche (CNEA), Instituto Balseiro (U. N. Cuyo), and
Conicet, 8400 Bariloche, R\'{\i}o Negro, Argentina.}
\date{\today }

\begin{abstract}
We have investigated the relationship between the domain structure and the
magnetic interactions in a series of FePt ferromagnetic thin films of
varying thickness. As-made films grow in the magnetically soft and
chemically disordered A1 phase that may have two distinct domain structures.
Above a critical thickness $d_{cr}\sim 30$ nm the presence of an out of
plane anisotropy induces the formation of stripes, while for $d<d_{cr}$
planar domains occur.

Magnetic interactions have been characterized using the well known DCD-IRM
remanence protocols, $\delta M$ plots, and magnetic viscosity measurements.
We have observed a strong correlation between the domain configuration and
the sign of the magnetic interactions. Planar domains are associated with
positive exchange-like interactions, while stripe domains have a strong
negative dipolar-like contribution. In this last case we have found a close
correlation between the interaction parameter and the surface dipolar energy
of the stripe domain structure. Using time dependent magnetic viscosity
measurements, we have also estimated an average activation volume for
magnetic reversal, $\langle V_{ac}\rangle \sim 1.37\times 10^{4}$ nm$^{3},$
which is approximately independent of the film thickness or the stripe
period.
\end{abstract}

\maketitle

\section{\label{introduction}Introduction}

\bigskip Ferromagnetic thin films exhibiting a magnetic domain structure in
the form of thin parallel stripes have been the subject of intense research
in the last few decades, both experimentally\cite%
{Saito,Hehn,Gehanno2,Okamoto,ButeraMFM,FeGa} and theoretically.\cite%
{Kooy,Murayama,Sukstanskii,Brucas,Hubert} This kind of structure is observed
in films that present an out of plane anisotropy component (due to stress,
crystalline texture, interfacial or other effects) and in a simplified
picture it can be described as a periodic pattern of parallel in-plane
magnetized regions in which the magnetization has a relatively small\
component that points alternatively in the two directions that are normal to
the film plane. A stripe (or bubble) pattern is generally observed for all
film thicknesses when the perpendicular anisotropy energy constant, $%
K_{\perp },$ is larger than the demagnetizing shape energy, $2\pi M_{s}^{2}$%
, ($M_{s}$ is the saturation magnetization) but can also be found below a
critical thickness $d_{cr}$ when $Q=K_{\perp }/2\pi M_{s}^{2}$ is smaller
than one. The transition from planar to stripe domains at $d=$ $d_{cr},$ is
due to the minimization of the total magnetic energy which can include the
contribution of anisotropy, demagnetizing, domain wall and Zeeman (for $%
H\neq 0$) terms. The critical thickness depends on the material properties
such as the effective anisotropy, the saturation magnetization and the
exchange stiffness constant, and also on the external field. There are
several models for the calculation of $d_{cr},$ see for example Refs. [%
\onlinecite{Sukstanskii,Brucas,Hubert}], that predict larger values of $%
d_{cr}$ in materials with a large saturation magnetization, a large
exchange, or a small anisotropy. The value of the critical thickness is in
the range of 20-30 nm for Co,\cite{Hehn} partially ordered FePd\cite%
{Gehanno2} or disordered FePt films\cite%
{ButeraMFM,Butera1,Butera2,ButeraFMR,ButeraSSW,Jonas,ButeraFMR2}, and can
take larger values (of the order of 200 nm) in films with lower anisotropy
such as permalloy.\cite{Ramos} Films with stripe domains have characteristic
$M$ vs. $H$ in-plane loops in which the following features are often
observed:\cite{Murayama,ButeraMFM} $i)$ the low field part of the curve
increases almost linearly from remanence until the saturation field is
reached. This in-plane saturation field was shown to increase with film
thickness following approximately the relationship $H_{\mathrm{sat}\parallel
}=H_{\mathrm{sat}\perp }\left[ 1-d_{cr}/\left( d\sqrt{1+Q}\right) \right] ,$
with $H_{\mathrm{sat}\perp }=2K_{\perp }/M_{s}.$ $ii)$ Due to the formation
of the stripe structure the in-plane coercivity increases abruptly and the
remanence decreases considerably above $d_{cr}.$ $iii)$ For $d\geq d_{cr}$
rotatable anisotropy,\textsl{\ i.e.} the alignment of the stripe structure
at remanence in the direction of a previously applied field, is observed.
The magnitude of this anisotropy also increases with film thickness and is
usually characterized by a field $H_{\mathrm{rot}}$. $iv)$ The period of the
stripe structure increases approximately as the square root of the film
thickness, $\lambda _{s}\varpropto \sqrt{d}.$

The study of the magnetic interactions present in films in which a crossover
from a planar to a striped magnetic domain structure is observed can then
give a deeper insight to understand this behavior. Both Henkel plots\cite%
{Spratt} and delta$-M$ ($\delta M)$ curves,\cite{Kelly,Otero} together with
the magnetic viscosity, $S,$\cite{Wohlfarth,Wohlfarth2} can be used to
estimate the sign of the magnetic interactions and the magnetic reversal
volumes in the samples. Magnetic interactions have been widely studied in
small particles,\cite{Che} thin continuous films,\cite{Che} granular systems%
\cite{CoAg} and nanostructured films\cite{NCA} using magnetic remanence
measurements.

The $\delta M$ curve is defined as the difference between two remanence
curves:%
\begin{equation}
\delta M=2M_{r}-1-M_{d},  \label{deltaM}
\end{equation}%
where the $M_{r}$ curve (also known as the isothermal remanent
magnetization, IRM) is obtained by starting from a state of zero remanence,
erased following a well defined protocol, and then measuring the
magnetization at zero field after applying fields of increasing magnitude.
The $M_{d}$ (or dc demagnetization, DCD) curve is obtained by saturating the
sample in a negative field and then repeating the same procedure as for the $%
M_{r}$ curves. These two curves are usually normalized to the remanence
saturation value ($M_{R}$) and labeled as $m_{r}$ and $m_{d}$. In the case
of a noninteracting system Wohlfarth predicted\cite{Wohlfarth2} that the two
remanence curves should be identical and hence $\delta M=0$. If $\delta
M\neq 0$ the effects of magnetic interactions can been accounted for using a
phenomenological model\cite{Che} for the effective interaction field, $%
h_{int},$ that takes into consideration dipolar-like (demagnetizing) and
exchange-like (magnetizing) interactions. In this model $h_{int}=\alpha
m+\beta (1-m^{2}),$ which means that the interaction field has a linear
dependence with $m$ (which can be both $m_{r}$ or $m_{d}$) with a slope of
magnitude $\alpha $. This parameter can be either positive or negative
depending on the dominant type of interaction, exchange-like or
dipolar-like, respectively. The term with the parameter $\beta $ accounts
for first order interaction field fluctuations from the mean field. A
numerical method to calculate $\alpha $ and $\beta $ is described in Ref. [%
\onlinecite{Che}], but they can be more easily obtained from the
experimental data following the procedure of Ref. [\onlinecite{Harrell}]%
\begin{equation}
\alpha =\int_{0}^{\infty }\delta M\,dh,\qquad \beta =\alpha /(3m_{r}^{0}-1).
\label{alpha}
\end{equation}%
In the above expression $h$ is the applied field normalized to the remanent
coercivity $H_{C}^{\mathrm{rem}}$ (defined as the reverse negative field
that, after saturation in the positive direction, produces a zero
magnetization at zero field) and $m_{r}^{0}$ is the remanent magnetization
at the point where $\delta M$ curves cross zero.

In order to get a deeper insight in the magnetic behavior, remanence
measurements are often complemented with magnetic relaxation experiments.
When a sample is magnetized in a negative saturating field and after that a
positive field is applied, the magnitude of $M$ often varies linearly with
the logarithm of time $t$. Changes in $M$ are due to thermally assisted
processes that provide the necessary energy to overcome the barrier energy
of magnitude $E$. The proportionality parameter is the magnetic viscosity $S$
and the relationship is often written as:\cite{Street}%
\begin{equation}
M(t,H)=M(t_{0},H)+S(H)\ln (t/t_{0}),  \label{viscosity}
\end{equation}%
with $t_{0}$ the initial time and $M(t_{0},H)$ the initial value of $M$ at $%
t=t_{0}$ for a given $H.$ The viscosity $S$ can be shown to depend on
temperature, $T$, saturation magnetization, $M_{s}$, and the distribution of
activation energies, $f(E)$, in the following way\cite{Gaunt1}%
\begin{equation}
S=2M_{s}k_{B}Tf(E).  \label{S}
\end{equation}%
The magnetic viscosity depends also on the forward applied field, through
the dependence of $f(E)$ on $H,$ and is generally maximum for an applied
field $H_{S}$ which is close to the macroscopic coercive field $H_{C}$.
Viscosity and remanence measurements can be related using the field
derivative of the DCD curve, known as the irreversible susceptibility\cite%
{Gaunt2}
\begin{equation}
\chi _{irr}=\frac{\partial M_{d}}{\partial H}=-2M_{s}f(E)\left( \frac{dE}{dH}%
\right) .  \label{Chi}
\end{equation}%
The variation of the activation energy with the magnetic field can be
related to the so called activation volume, $\left\vert \frac{dE}{dH}%
\right\vert =cV_{ac}M_{s}$, where $c$ is a constant of the order of unity
and its value depends on the kind of system that is under consideration.
Simple calculations\cite{Gaunt2} for monodomain particles or strong
domain-wall pinning give $c=1,$ while for weak domain-wall pinning $c=2.$ If
demagnetizing effects are considered,\cite{Ng} $c=4$ for strong pinning and $%
c\geq 2$ for weak pinning. Using Eqs. [\ref{S}] and [\ref{Chi}] the
activation volume can be written as:%
\begin{equation}
V_{ac}=\frac{k_{B}T\chi _{irr}}{cM_{s}S}.  \label{Vac}
\end{equation}%
In the case of thin films in which the magnetization changes by a process of
domain wall motion, the activation volume can be interpreted as the volume
swept by a single jump between pinning centers. This volume is usually
related with the fluctuation field, $H_{f}$, defined as:\cite{Ng}%
\begin{equation}
H_{f}=\frac{S}{\chi _{irr}}=-\frac{k_{B}T}{dE/dH}=\frac{k_{B}T}{cM_{s}}\frac{%
1}{V_{ac}}.  \label{Hf}
\end{equation}

Magnetic interactions in FePt have been investigated in different systems,
including continuous films,\cite{Li,Jeong} annealed multilayers,\cite{Luo}
exchange-coupled bilayers,\cite{Pernechele} granular films,\cite{Wei,Wei2}
and nanoparticles,\cite{Gao} all in the atomically ordered L1$_{0}$ phase.
Negative interparticle interactions were reported in the cases of films,
annealed multilayers and nanoparticles, when the external field was applied
parallel to the in-plane direction (these films show in-plane anisotropy).
On the other hand, continuous films exhibiting out of plane anisotropy\cite%
{Jeong,Wei} present positive $\delta M$ curves when remanence curves are
measured perpendicular to the film plane. Magnetic relaxation has been
reported in the case of exchange-coupled Fe/FePt bilayers,\cite{Pernechele}
annealed Fe/Pt multilayers,\cite{Luo} and polycrystalline thin films\cite%
{Jeong} all of them in the hard magnetic phase. For a single layer of 10 nm
of FePt with an average grain size of $\sim 20$ nm, the authors in Ref. [%
\onlinecite{Pernechele}] reported $V_{ac}\sim $ 12500 nm$^{3}.$ In the
second case the authors estimated $V_{ac}\sim $ 1200 nm$^{3}$ for a
multilayer with a total thickness of 15 nm. In the last case an activation
volume $V_{ac}\sim $ 400 nm$^{3}$ was estimated for a film 5 nm thick with a
crystalline grain size of 10 nm. This last sample presented a maze structure
of magnetic domains at remanence, consisting of irregular elongated regions
magnetized perpendicular to the film plane with a length of several
micrometers and a width of 100-150 nm. Assuming spherical reversal volumes,
the corresponding "activation diameters" are $d_{ac}=29$, 13, and 9 nm,
respectively. Note that if the activation volume is divided by the film
thickness, and cylindrical domains are assumed, the resulting "activation
length" is in the range of 40 nm for the first sample and 9 nm in the last
two systems.

As far as we know, magnetic interactions and time dependent effects in FePt
films in the A1 disordered phase have not been yet characterized. The
possibility to tune the domain structure by varying the film thickness can
be used to study how these effects are affected by the way in which the
magnetic domains order. In the following sections we present a detailed
experimental study in a series of as-made FePt thin films of different
thicknesses in which the magnetic interactions have been investigated by
means of DCD-IRM, delta-$M$ plots and magnetic viscosity measurements.

\bigskip

\section{\label{ExpDetails}Experimental details}

FePt films have been fabricated by dc magnetron sputtering on naturally
oxidized Si (100) substrates. A detailed description of the preparation and
the structural characterization can be found in Ref. [\onlinecite{ButeraMFM}%
]. The samples were deposited from an FePt alloy target with a
nominal atomic composition of 50/50. We sputtered eight films with
thicknesses of\ 9, 19, 28, 35, 42, 49, 56 and 94 nm. The samples
were studied using X-Ray diffraction, transmission electron
microscopy (TEM) and energy-dispersive X-Ray spectroscopy (EDS)
techniques. The X-ray diffractograms showed that the samples grow in
the fcc A1 crystalline phase, without traces of the ordered L1$_{0}$
structure. A [111] texture normal to the film plane was observed and
comparison with stress released films revealed that as-made samples
were also subjected to an in-plane compressive stress. An average
crystallite grain diameter of 4 nm was obtained from TEM
micrographs. The photoemission spectra indicated that the Fe/Pt
atomic ratio of the films was approximately $45$/55. Stress effects
are the main contribution to an
effective magnetic anisotropy perpendicular to the film plane of magnitude $%
K_{\perp }=1.5(4)\times \ 10^{6}$ erg/cm$^{3}$, which gives rise to
a magnetic domain structure in the form of stripes for $d>d_{cr}\sim
30$ nm. As we have already shown in Ref. [\onlinecite{ButeraMFM}]
using magnetic force microscopy (MFM) techniques, the half period of
the stripe pattern scales with the square root of the film thickness
starting at $\lambda /2\sim 45$ nm for $d=35$ nm and reaching
$\lambda /2\sim 75$ nm for $d=94$ nm. For $d<d_{cr}$ an in-plane
planar domain structure is observed. In both domain regimes a strong
correlation between the domain configuration and the shape of the
hysteresis loops was found.

The DC demagnetization (DCD), Isothermal Remanent Magnetization (IRM) and
viscosity data were measured using a LakeShore model 7300 VSM, capable of a
maximum field of 10000 Oe. For the DCD measurements we used the following
sequence of applied fields $(-H_{sat},\Delta H,0);(-H_{sat},2\Delta
H,0);...(-H_{sat},n\Delta H,0);...$ In this case a negative saturation field
$-H_{sat}$ is applied before each data point is acquired at $H=0$ after
applying a field $H=n\Delta H.$ In most cases we set $H_{sat}=5000$ Oe and $%
\Delta H\leq 10$ Oe, depending on the coercivity of the sample. A waiting
time of 5 seconds was used before measuring the remanent magnetization.
There is an alternative field sequence for performing DCD experiements\cite%
{Wang} $-H_{sat};(\Delta H,0);(2\Delta H,0);...$ in which the saturation
field is applied only at the beginning of the experiment. In principle, this
method should be less sensitive to the waiting time and the field step $%
\Delta H$, and differences between the DCD and IRM curves due to viscosity
effects are minimized. In our case we did not observe significant
differences between both DCD sequences and decided to use the first method.

The IRM curve is obtained by starting from a demagnetized state and
measuring the magnetization at zero field following the sequence $(\Delta
H,0);(2\Delta H,0);...$The ideal demagnetized remanent state is the one
obtained by heating the sample above the Curie temperature, $T_{C}$, and
then cooling in zero field. Because of the appearance of irreversible
effects in the magnetic response,\cite{Jonas} our films can not be heated to
$T_{C}$ $\sim 500$ K, so we adopted two different protocols to demagnetize
the samples. The "linear" demagnetization routine is the usual procedure in
which the sample is saturated in one direction and a sequence of decreasing
fields is applied in both senses, until zero field is reached. Films can be
also demagnetized in a slowly decreasing field (from saturation to zero)
while they are quickly rotated around an axis perpendicular to the magnetic
field. The "rotating" demagnetization routine usually gives a remanent state
that is more disordered and isotropic in the film plane than in the linear
case, resembling the state that can be obtained by cooling the sample from
above $T_{C}$.

In the case of magnetic relaxation measurements films were saturated in a
negative field of 5000 Oe, a positive field was then applied and kept
constant during the whole experiment while the magnetization was measured in
intervals of 10 seconds during approximately 30 minutes. We calculated the
viscosity from the linear fit of the time variation of $M$ (Eq. \ref%
{viscosity}). The same routine was repeated for several fields in the
vicinity of $H_{C}$ from which the magnetic viscosity $S(H)$ is obtained.

\section{\label{ExpResults}Experimental results and discussion}

\subsection{IRM and DCD\ measurements}

\begin{figure}[tbph]
\includegraphics[ width=9cm]{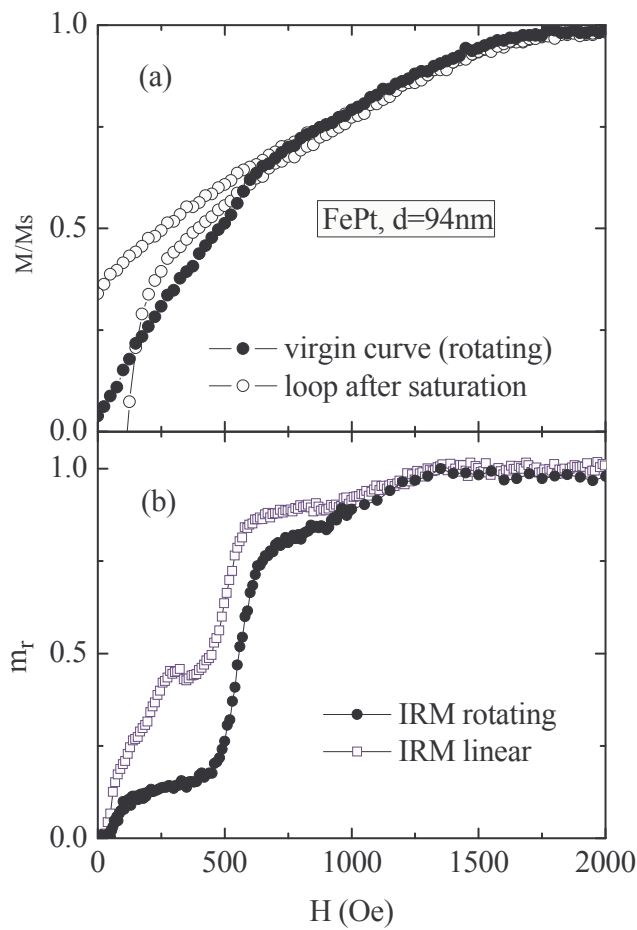}
\caption{(color online) a) Upper right quadrant of the normalized hysteresis
loop for $d=94 $ nm. The virgin curve was obtained after demagnetizing the
film with the "rotating" routine. Note that there is a range of fields in
which the virgin curve stays outside of the hysteresis loop. b) Normalized
IRM\ curves obtained in the same sample after it was demagnetized using the
"linear" or the "rotating" routines. The coercive field for this film is $%
H_{C}=125$ Oe.}
\label{FigDemag}
\end{figure}

In all films we have measured the IRM curves using the two demagnetizing
sequences mentioned in the previous section. For films with $d\leq 19$ nm
additional care must be taken in order to reach a truly demagnetized state,
because the magnetization switching at $H=H_{C}$ occurs in a very narrow
field range of only a few Oe. The differences between "rotating" and
"linear" demagnetizing routines are more pronounced in thicker films. In
Fig. \ref{FigDemag} (a)\ we show the upper right quadrant of the hysteresis
loop for the film with $d=94$ nm, together with the virgin curve obtained
after demagnetizing the film using the rotating routine. It can be observed
that there is a field region in which the virgin curve is not within the
hysteresis loop. This effect is almost absent when the sample is
demagnetized using the linear sequence and to explain it one must consider
that the remanent state obtained when the sample is demagnetized using the
rotating routine consists of an array of randomly oriented stripes.\cite%
{ButeraMFM} On the other hand, in the case of the linear protocol almost all
stripes are already aligned at remanence in the direction of the
demagnetizing field. When the sample is saturated, rotational anisotropy
imposes an easy magnetization axis along the field direction and the stripes
are always aligned in that direction. Taking into account these effects one
can understand why in the linear case the virgin curve stays inside the
loop, while after the rotating cycle larger fields are needed to reach the
same magnetization value because part of the field energy is used in
aligning the stripes in the direction of the applied field.

The same differences are observed in the IRM curves, as can be seen in Fig. %
\ref{FigDemag} (b). In this case starting from an initially more disordered
and isotropic state (rotating routine), makes more difficult the
magnetization of the sample in the direction of the applied field. Note that
the magnetization process occurs in several steps. In the low field region,
a relatively fast initial increase of $m_{r}$ (from $m_{r}=0$ to $m_{r}\sim
0.10)$ occurs for fields $H\lesssim H_{C}$ $\sim 125$ Oe, for $d=94$ nm. We
associate these changes to domain wall movement in the small fraction of
regions which were already aligned in the direction of the applied field.
Then $m_{r}$ stays relatively constant until $H\sim 500$ Oe, which is more
or less coincident with a kink in the virgin magnetization curve or the
beginning of the reversible part of the $M-H$ loop. These features were
assigned in Ref. [\onlinecite{ButeraMFM}] to the rotational anisotropy field
$H_{\mathrm{rot}}$, the field necessary to rotate the in-plane easy axis of
the stripes in the direction of the applied field. Once the stripes are
aligned they can be more easily moved by the mechanism of domain wall
displacement and a very large increase in $m_{r}$ (from $m_{r}=0.15$ to $%
m_{r}=0.75)$ occurs in the range $H=500-650$ Oe. Comparison with the
hysteresis loop suggests that for $H>650$ Oe and until $H\sim 1400$ Oe
irreversible changes in $m_{r}$ are probably due to the rotation of regions
that are magnetized perpendicular to the film plane. The linearly
demagnetized IRM curve shows similar characteristics, but the irreversible
changes at low fields ($H\sim H_{C}$) are considerably larger, with $m_{r}$
reaching almost 50\% of the saturation value. Above this field a rapid
increase and then a more gradual approach to saturation is observed, with
the same overall behavior already described for the rotating routine. In the
rest of the films there are still differences between both demagnetizing
protocols, but they tend to disappear as the films become thinner. For $%
d\leq 35$ nm both IRM curves are almost identical.
\begin{figure}[tbph]
\includegraphics[ width=10cm]{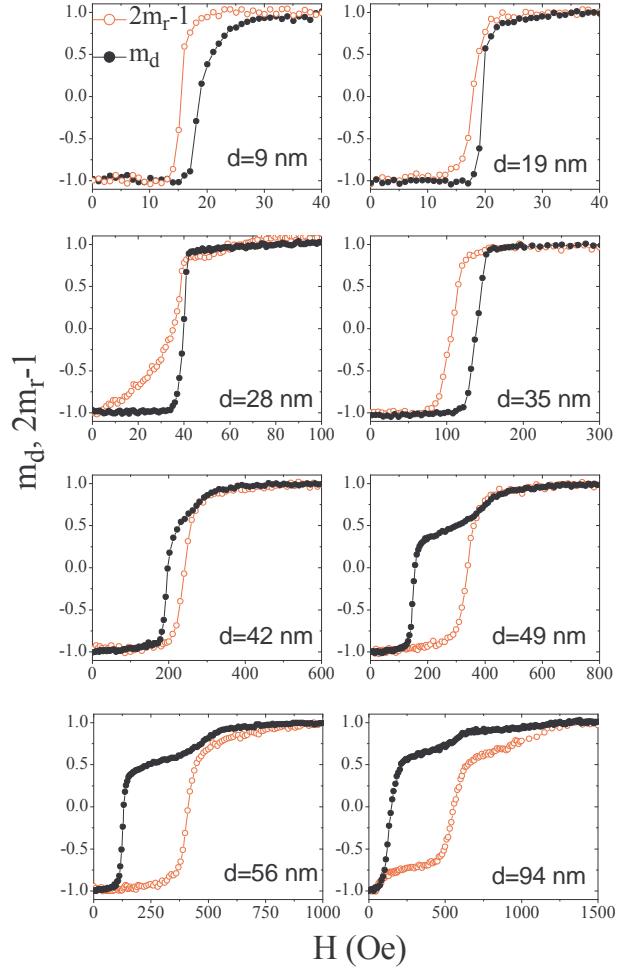}
\caption{(color online) Normalized IRM and DCD remanence curves for the
different films. In all cases the IRM data were acquired using the rotating
protocol to obtain the demagnetized state. }
\label{FigIRM-DCD}
\end{figure}

In Fig. \ref{FigIRM-DCD} we show the normalized DCD and IRM (starting from a
rotating demagnetizing cycle) curves for all films. We have plotted $m_{d}$
and $2m_{r}-1$ in order to compare both measurements. The most significant
feature that can be observed is that for $d\leq 35$ nm the IRM is above the
DCD curve and the relationship is inverted for $d\geq 42$ nm. As can be
deduced from Eq. \ref{deltaM} this implies a change in sign in the $\delta M$
curve that is indicating a change in the dominant magnetic interactions. The
fact that $2m_{r}-1>$ $m_{d}$ in the case of thinner films is telling us
that in these samples the saturated state can be reached more easily, i.e.
the magnetic interactions favor a magnetized state. In thicker films the IRM
is always below the DCD\ curve, which reflects that dipolar-like
interactions are dominant in these samples.

\begin{figure}[tbph]
\includegraphics[ width=8cm]{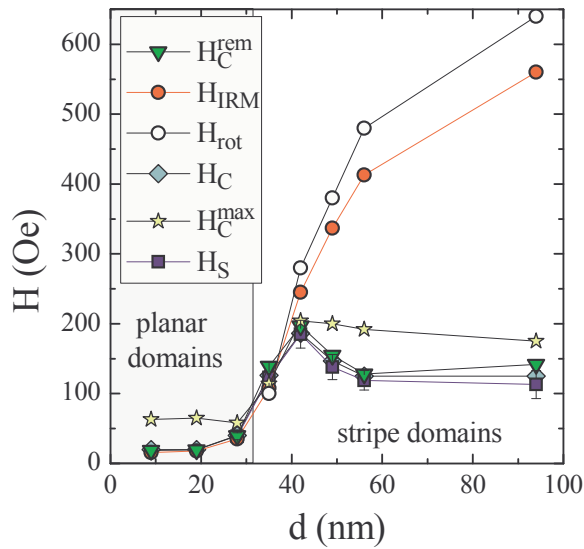}
\caption{(color online) Room temperature remanence fields $H_{C}^{\mathrm{rem%
}}$ and $H_{\mathrm{IRM}}$ obtained from the zero crossing of the $m_{d}$
and $2m_{r}-1$ curves, respectively. We have also plotted for comparison the
room temperature coercive field, $H_{C}$, and $H_{S},$ the field where the
maximum in the magnetic viscosity is found. $H_{\mathrm{rot}}$ has been
extracted from Ref. [\onlinecite{ButeraMFM}] and is a measure of the average
field needed to rotate the stripe structure by 90$^{\circ }$. $H_{C}^{\max }$
is the maximum value of $H_{C}$ in the temperature interval 4-300 K (taken
from Ref. [\onlinecite{Jonas}]).}
\label{FigHrem}
\end{figure}

From the field where $m_{d}$ and $2m_{r}-1$ curves cross zero we can extract
the remanent coercivity $H_{C}^{\mathrm{rem}}$ and the IRM half reversal
field, $H_{\mathrm{IRM}}$, respectively. We will show later that the
normalized difference $2\left( H_{C}^{\mathrm{rem}}-H_{\mathrm{IRM}}\right)
/H_{C}^{\mathrm{rem}}$ may be used as a very good estimation of the sign and
magnitude of the magnetic interactions. This quantity is very similar to the
so-called Interaction Field Factor, IFF$=\left( H_{C}^{\mathrm{rem}}-H_{%
\mathrm{IRM}}\right) /H_{C}$, differing only in the normalization variable.
In Fig. \ref{FigHrem} we plotted these two fields, together with the
coercivity $H_{C}$, and the field $H_{\mathrm{rot}}$ obtained from Ref. [%
\onlinecite{ButeraMFM}]. This field is a measure of the average magnetic
field needed to overcome the rotational anisotropy. For a Stoner-Wohlfarth
system the two remanence fields should have the same value as $H_{C},$ which
is relatively small for the thinner films, increases considerably when the
stripe structure is formed, has a maximum at $d\sim 42$ nm and levels off at
$H_{C}\sim 140$ Oe for larger thicknesses. This behavior is approximately
followed by $H_{C}^{\mathrm{rem}},$ although as expected $H_{C}^{\mathrm{rem}%
}>H_{C}$, but is definitely not true for $H_{\mathrm{IRM}}$. The IRM\
reversal field increases continuously with film thickness giving another
indication of the change in the magnetic interactions when the stripe
structure is formed. As already discussed in the case of the $d=94$ nm film
the IRM curve is a fingerprint of the field necessary for gradually aligning
the domains that are not parallel to $H$ in the direction of the applied
field. It is then expected that $H_{\mathrm{IRM}}$ values follow closely the
thickness dependence of $H_{\mathrm{rot}}$, the field needed to overcome the
rotational anisotropy. As can be seen in Fig. \ref{FigHrem} both fields
follow a similar trend, with the differences in the absolute values arising
from the different remagnetizing mechanism that $H_{\mathrm{IRM}}$ and $H_{%
\mathrm{rot}}$ describe.

\begin{figure}[tbph]
\includegraphics[ width=10cm]{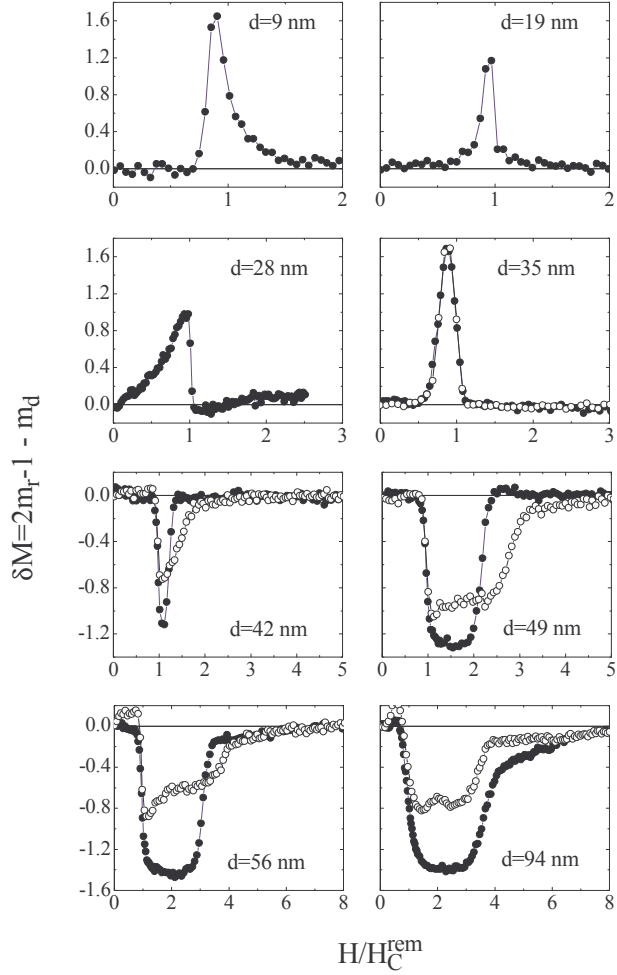}
\caption{Delta-$M$ plots for all the studied samples that indicate the
deviation from a Stoner-Wohlfarth behavior. For the thicker films we have
plotted the data obtained using the two demagnetization protocols for the
IRM curves with full and open symbols (rotating and linear demagnetization
routines, respectively). }
\label{FigdeltaM}
\end{figure}

One of the methods to characterize qualitatively the magnetic interactions
is by using the $\delta M$ plots (see Eq. \ref{deltaM}), which reflect the
deviations from the Stoner-Wohlfarth behavior. As already mentioned, if the
IRM is above the DCD curve the $\delta M$ plot is positive and the
interactions tend to be of the exchange type, favoring a magnetized state.
Dipolar-like interactions are more important when $\delta M$ is negative. In
Fig. \ref{FigdeltaM} we show the $\delta M$ plots for all the studied
samples as a function of the applied field (normalized by the remanent
coercivity, $H_{C}^{\mathrm{rem}}$). We have used full symbols to indicate $%
\delta M$ plots obtained from an IRM\ curve that was isotropically
demagnetized (rotational routine) and open symbols for the case of a
linearly demagnetized sample. Note that there are differences between both $%
\delta M$ curves in the case of thicker films that tend to decrease
gradually as the thickness is decreased. For $d\leq 35$ nm the two
curves are almost coincident. These results show again explicitly
that the dominant interaction changes from magnetizing to
demagnetizing when the stripe structure starts to develop at $d\sim
35$ nm and they also give additional evidence of the effects of the
rotational anisotropy on the IRM remanence curves. We have already
shown in Figs. \ref{FigIRM-DCD} and \ref{FigHrem} that the fields
where the DCD and IRM curves cross zero are more separated in the
case of thicker films. This difference explains the shift in the
minimum in the $\delta M$ plots from $H/H_{C}^{\mathrm{rem}}=1$ to
at least twice this value for $d=94$ nm.

\begin{figure}[tbph]
\includegraphics[ width=8cm]{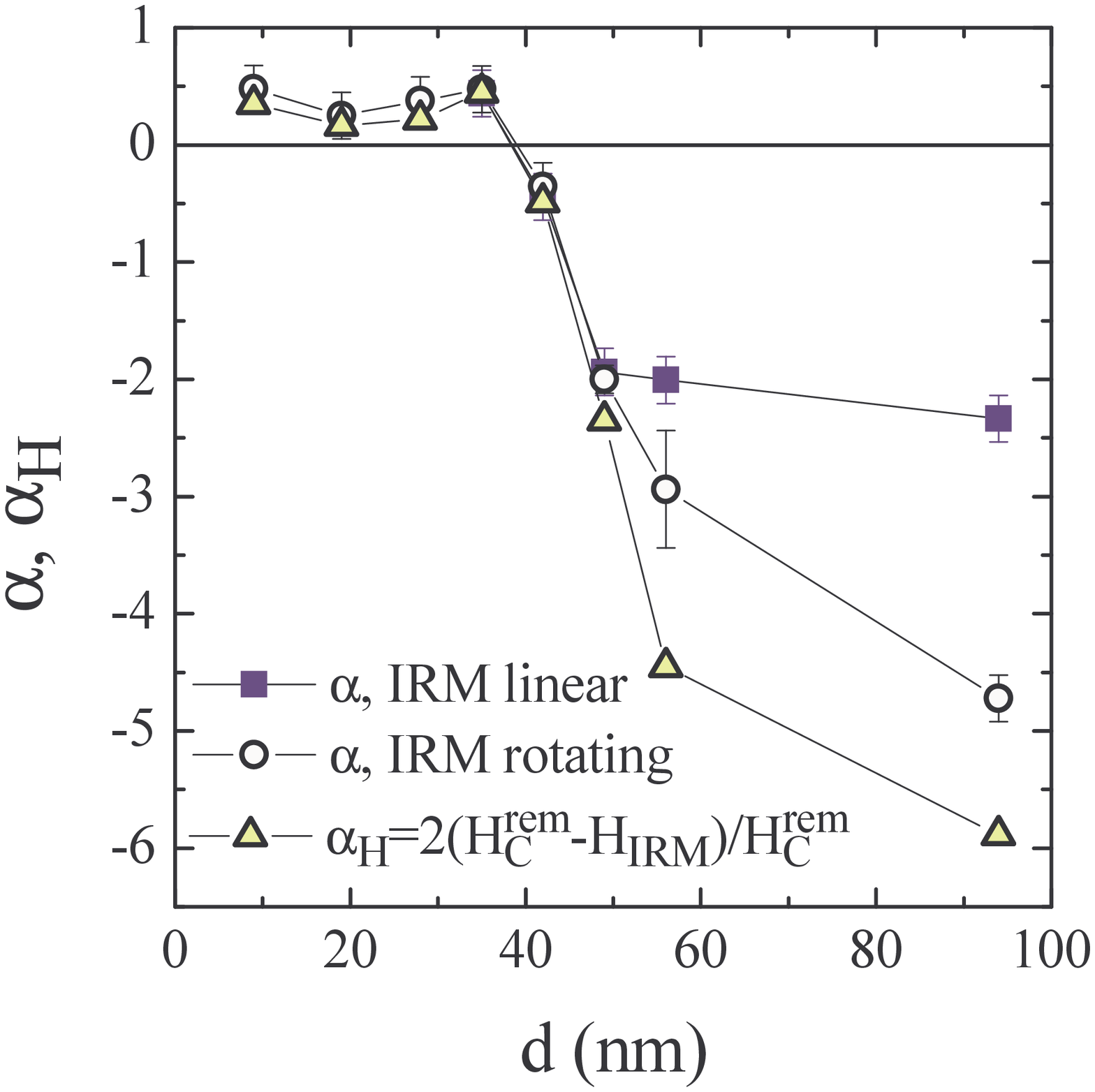}
\caption{(color online) Interaction parameter $\protect\alpha $ as a
function of film thickness obtained from the integration of the $\protect%
\delta M$ plots. Open symbols correspond to $\protect\alpha $ values
obtained from isotropically demagnetized (rotating routine) IRM curves while
full symbol data were obtained from linearly demagnetized samples. The
magnitude $\protect\alpha _{H}=2(H_{C}^{\mathrm{rem}}-H_{\mathrm{IRM}%
})/H_{C}^{\mathrm{rem}}$ is plotted for comparison and it is found to be
quite similar to $\protect\alpha $.}
\label{Figalpha}
\end{figure}

An estimation of the strength of the magnetic interactions can be obtained
from Eq. \ref{alpha}, which gives the interaction parameter $\alpha $ of the
Che and Bertram model.\cite{Che,Harrell} The integral of the $\delta M$
plots as a function of $d$ is presented in Fig. \ref{Figalpha}. Again we
show values of $\alpha $ obtained with both demagnetizing routines. We have
plotted in the same figure the quantity $\alpha _{H}=2\left( H_{C}^{\mathrm{%
rem}}-H_{\mathrm{IRM}}\right) /H_{C}^{\mathrm{rem}}$ which may be also used
to estimate the magnetic interactions. In the case of perfectly square $m_{d}
$ and $2m_{r}-1$ curves, the values of $\alpha $ and $\alpha _{H}$ should be
the same because the $\delta M$ plot is rectangular with an area $2\left(
H_{C}^{\mathrm{rem}}-H_{\mathrm{IRM}}\right) .$ Due to the different
distribution of switching fields in the IRM and DCD curves, the values of $%
\alpha $ differ from this simple estimation but, as can be observed in Fig. %
\ref{Figalpha}, the values and the shape of the curves of $\alpha $ and $%
\alpha _{H}$ as a function of film thickness are very similar, confirming
that $\alpha _{H}$ is also a very reasonable parameter for the estimation of
the magnetic interactions.

\begin{figure}[tbph]
\includegraphics[ width=8cm]{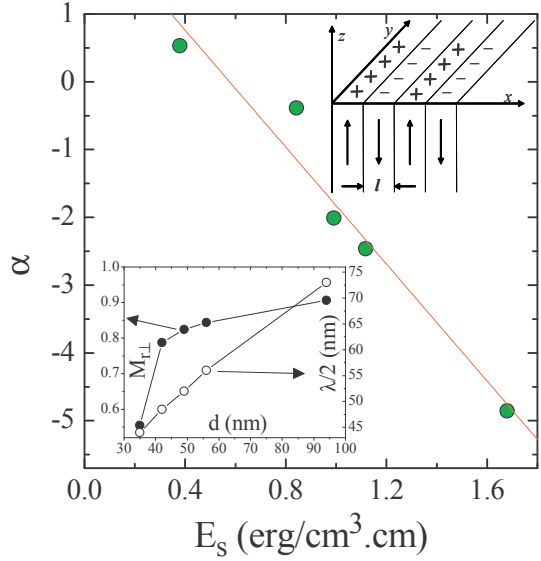}
\caption{(color online) Dependence of the interaction parameter $\protect%
\alpha $ (obtained using the rotating routine) as a function of the surface
magnetostatic energy of the domain configuration sketched in the inset. We
also show the dependence of the perpendicular remanence and the stripe half
period, taken from Ref. [\onlinecite{ButeraMFM}].}
\label{Fig6}
\end{figure}

A dimensional analysis of Eq. \ref{alpha} reveals that the interaction
parameter $\alpha $ may be associated to a normalized energy and hence could
be correlated with the dominant energy contribution to the magnetic domain
configuration. In the case of domains formed by parallel slabs of size $l$
magnetized perpendicular to the film plane (see the sketch in Fig. \ref{Fig6}%
) it is possible to calculate\cite{Chikazumi} the magnetostatic energy per
unit surface area as $E_{S}[\mathrm{erg}/\mathrm{cm}^{2}]=0.374M_{\perp
}^{2}l.$ We can estimate this energy for the different films that show a
stripe structure by identifying the thickness of the slabs with the values
of the half period of the stripe structure ($l=\lambda /2),$ and estimating
the component of the magnetization perpendicular to the film plane as $%
M_{\perp }(d)/M_{s}=M_{r\perp }(d)/M_{s}\sim \sqrt{1-[M_{r\parallel
}(d)/M_{r\parallel }(d=28)]^{2}}.$ The thickness dependence of $\lambda /2$
and $M_{r\perp }$ is shown in the inset of \ref{Fig6}. In the last formula $%
M_{r\parallel }(d)$ is the remanence in the direction of the applied field
(obtained from the saturation value of DCD or IRM measurements) and was
normalized by $M_{r\parallel }(d=28$ nm$)$ instead of $M_{s}$ to consider
that there is always a small component of $M$ that is neither parallel to
the anisotropy axis induced by $H$ nor parallel to the film normal. In the
main panel of Fig. \ref{Fig6} we plotted the interaction parameter $\alpha $
as a function of $E_{S}$ for $d\geq 35$ nm and found that there is a good
linear correlation between both magnitudes, indicating that for films with $%
d\geq d_{cr}$ it is energetically favorable to form a stripe structure which
has a magnetostatic energy that increases with the stripe period (and the
film thickness).

\subsection{\protect\bigskip Magnetic viscosity\ measurements}

\begin{figure}[tbph]
\includegraphics[ width=9cm]{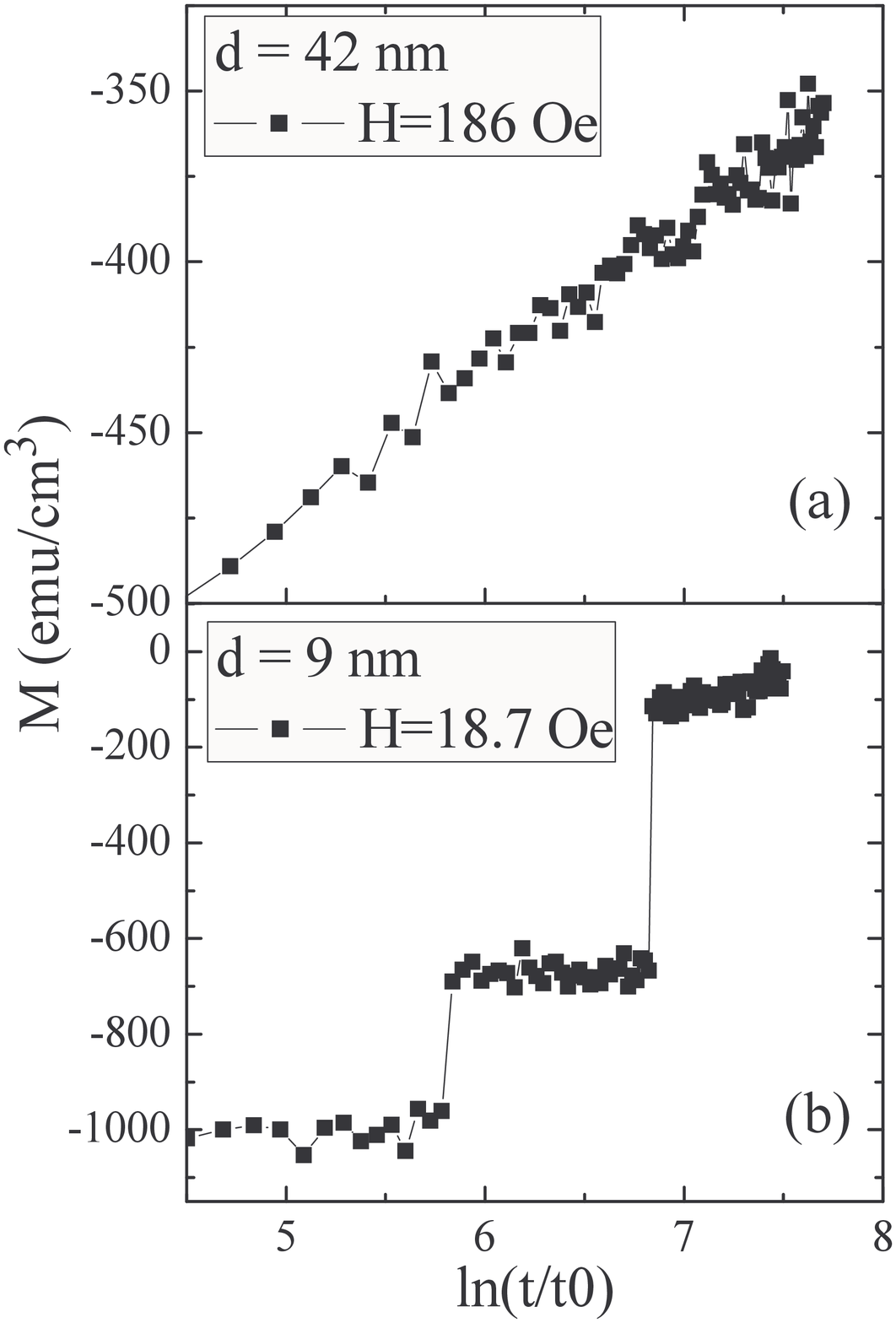}
\caption{Magnetic relaxation in films of different thickness close to the
coercive field. In panel a)\ we show the behavior of a film of 42 nm, which
closely obeys a logarithm law. Panel b)\ corresponds to a film of 9 nm that
presents a discontinuous relaxation.}
\label{FigRelaxation}
\end{figure}

Magnetic relaxation measurements were also performed in the whole set of
samples. For films with $d>28$ nm we found that Eq. \ref{viscosity} is
closely obeyed (see Fig. \ref{FigRelaxation} (a)) while in the case of
thinner films (9 nm and 19 nm) the relaxation of the magnetization follows a
nonlogarithm behavior or occurs in discrete steps, as can be observed in
Fig. \ref{FigRelaxation} (b). This last behavior has been only detected for
fields very close to $H_{C}$ and is an indication of the very narrow
distribution of energy barriers (or switching field distribution) in the
thinner films. For the relaxation measurements in these samples we took data
every 0.2 Oe which is almost equal to the stability limit (0.1 Oe) of the
electromagnet power supply. Possible fluctuations in the applied field can
switch the magnetization and it is then difficult to conclude that in this
case the reversal of the magnetization is only due to thermal effects. The
film with $d=28$ nm was at the limit where a reasonably linear fit could be
obtained and was included in the viscosity data, although with a larger
uncertainty in the determination of $S$.
\begin{figure}[tbph]
\includegraphics[ width=9cm]{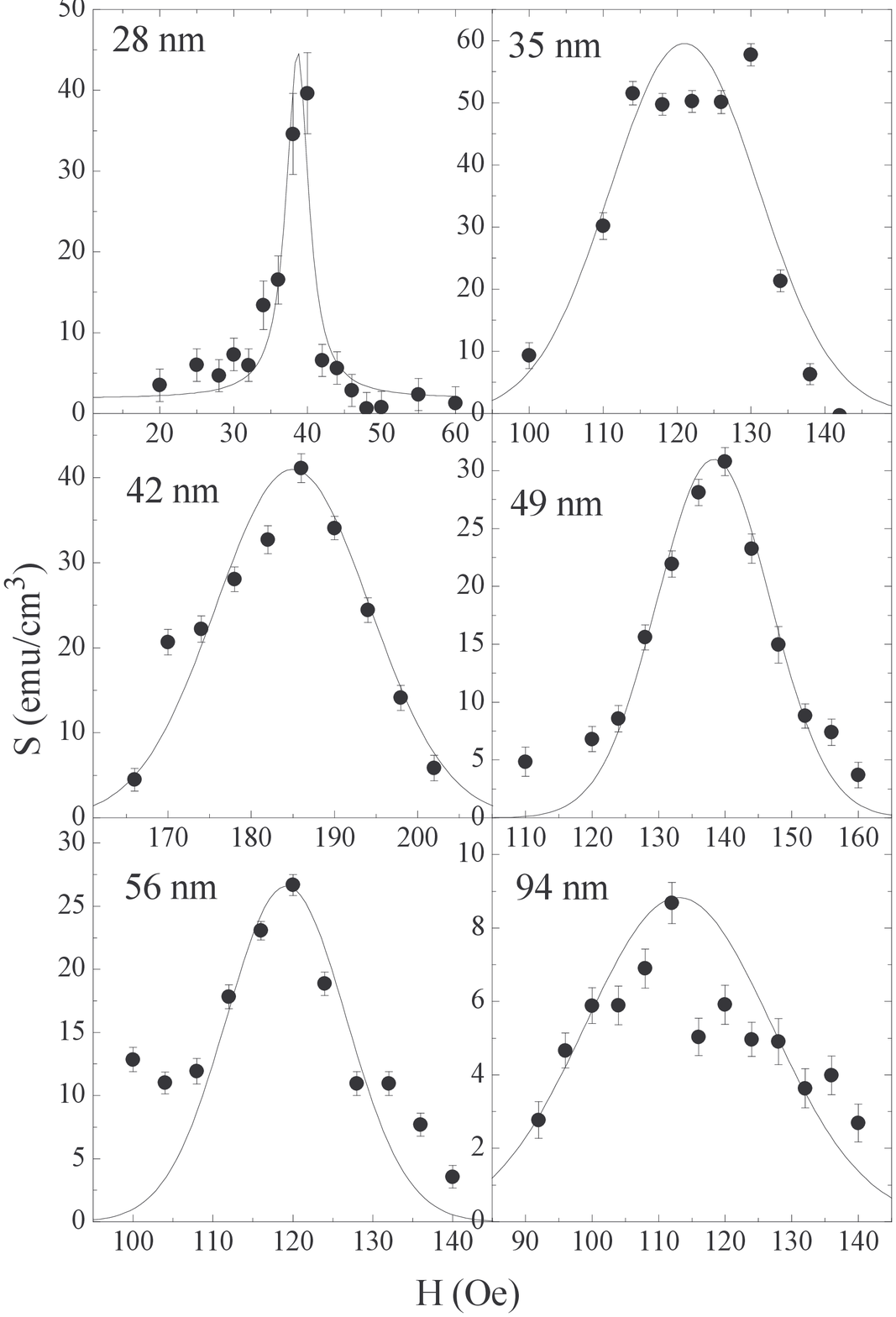}
\caption{Magnetic viscosity as a function of field in the vicinity of the
coercive field. Data are presented for the different samples in which a
reasonably linear variation of magnetization with $\ln (t/t_{0})$ was
observed. Different sets of data have been fitted with a gaussian
distribution from which we extracted $S_{\max }$, $H_{S}$ and $\Delta H_{S}$%
. }
\label{FigSvsH}
\end{figure}

The viscosity parameter, obtained from the slope of curves similar to Fig. %
\ref{FigRelaxation} (a), is plotted in Fig. \ref{FigSvsH} for the different
films as a function of the applied field. In all cases we observed a maximum
value of viscosity, $S_{\max }$, at a field $H_{S}$ which is close, but
always smaller, than $H_{C}$ (see Fig. \ref{FigHrem})$.$ The distribution of
viscosity values around $S_{\max }$ has a field width at half maximum height
(FWHM) characterized by $\Delta H_{S}$ which is very narrow for $d=28$ nm ($%
\Delta H_{S}\sim 3$ Oe), increases to an average value $\Delta H_{S}\sim 20$
Oe for $35\leq d\leq 56$ nm and increases again to $\Delta H_{S}\sim 60$ Oe
for $d=94$ nm. As already discussed in Section \ref{introduction}, the field
dependence of $S$ is a measure of the distribution of energy barriers (see
Eq. \ref{S}) and should correlate closely with the irreversible
susceptibility obtained from the derivative of the DCD curves.

\begin{figure}[tbph]
\includegraphics[ width=9cm]{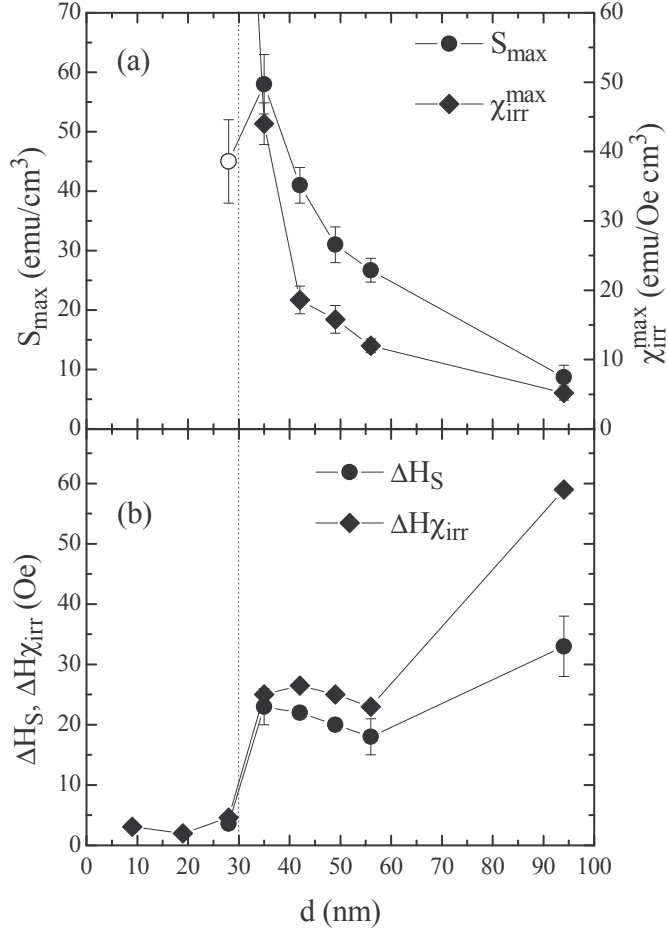}
\caption{(a) Maximum of the magnetic viscosity and the irreversible
susceptibility as a function of film thickness. The open symbol for
$d=28$ nm indicates that this sample is at the boundary in which a
reasonably linear behavior of $M(t)$ is observed. (b) Field
distribution width of both parameters as a function of $d$ (obtained
from a gaussian fit of the curves of Fig. \protect\ref{FigSvsH} and
the field derivative of the $m_{d}$ data of Fig.
\protect\ref{FigIRM-DCD}). The dotted vertical line indicates the
value of $d_{cr} \sim 30$ nm. } \label{FigSmax}
\end{figure}

In Fig. \ref{FigSmax} (a) we present the thickness dependence of the maxima
in the magnetic viscosity and the irreversible susceptibility, $S_{\max }$
and $\chi _{irr}^{\max }$, obtained from Figs. \ref{FigSvsH} and \ref%
{FigIRM-DCD}, respectively, and in the lower panel of the same figure we can
observe the FWHM value of the field distribution of both magnitudes. As
expected, the same overall behavior of $S_{\max }$ and $\chi _{irr}^{\max }$
is found for all samples with the exception of $d=28$ nm which has been
indicated with an open symbol in Fig. \ref{FigSmax} (a). As we already
mentioned this film is at the limit in which a logarithm time decay of $M$
is found and, as can be seen in Fig. \ref{FigSvsH}, it has a very narrow
field distribution which complicates the precise determination of $S_{\max }$%
. It is then quite possible that the real value of the maximum viscosity for
$d=28$ nm be considerably larger than the reported value, that should then
be considered as a lower limit of $S_{\max }.$ Discarding this value of
viscosity, it is observed that $S_{\max }$ decreases with film thickness,
indicating that the magnetic relaxation in thinner films is faster than in
thicker samples. As expected from Eqs. \ref{S} and \ref{Chi} and observed in
Fig. \ref{FigSmax} (b) the field distribution of both $S_{\max }$ and $\chi
_{irr}^{\max }$ has the same thickness dependence, which indicates that the
distribution of activation energies $f(E)$ tends to be considerably narrower
for films with $d<d_{cr}$. The sharpness of $\chi _{irr}$ peaks (\textit{i.e.%
} smaller $\Delta H\chi _{irr}$ values) has been argued\cite{Tomka} to be an
indication of strong exchange interactions between neighbor grains,
consistent with our findings from $\delta M$ curves.

\subsection{\protect\bigskip Activation volume and fluctuation field}

\begin{figure}[tbph]
\includegraphics[ width=9cm]{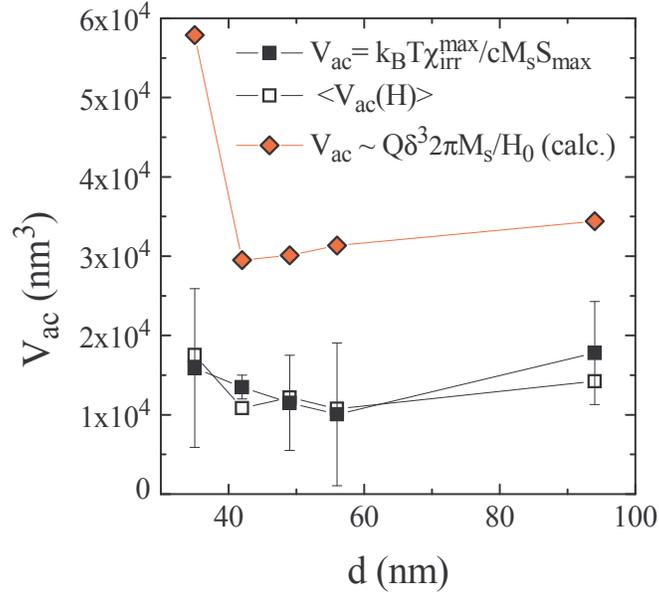}
\caption{(color online) Activation volume as a function of film thickness.
We have calculated $V_{ac}$ from the ratio $\protect\chi _{irr}^{\max
}/S_{\max }$ (open squares) and from the average values obtained from
measurements at different fields in the vicinity of $H_{C}$ (full squares).
The diamond symbols correspond to the calculation of the activation volumes
using the model of Gaunt.\protect\cite{Gaunt1}}
\label{FigVac}
\end{figure}

The activation volume can be calculated from Eq. \ref{Vac} using the ratio
between the maximum values of $S$ and $\chi _{irr}$ or by averaging
different values of $V_{ac}(H)$ in the vicinity of the coercive field. To
estimate the parameter $c$ entering in Eq. \ref{Vac} we need to know the
reversal mechanism present in our films. We have measured the out of plane
angular variation of the coercive field and found that $H_{C}$ increases
when the field is applied at increasing angles with respect to the film
plane, an indication that reversal is due to the displacement of domain
walls. For this case there is a criterion given by Gaunt\cite{Gaunt1} for
the determination of the pinning regime. He defined a parameter $\beta
_{0}=3F/(2\pi \gamma \delta ),$ where $F$ is the maximum restoring force a
pin can exert on a wall, $\gamma $ is the wall energy and $\delta $ the wall
width. For $\beta _{0}<1$ the domain walls are in the weak pinning regime
while for $\beta _{0}>1$ the strong pinning situation occurs. A crude
estimation for the pinning force is given by $F=1/2(4\pi M_{s}a/3)^{2}$ ($a$
is the radius of the pinning centers or inclusions) and the wall energy can
be written as $\gamma =K\delta ,$ so that we can write:
\begin{equation}
\beta _{0}=\frac{4\pi M_{s}^{2}}{3K}\left( \frac{a}{\delta }\right) ^{2}=%
\frac{2}{3Q}\left( \frac{a}{\delta }\right) ^{2}.  \label{beta}
\end{equation}%
In our films we have\cite{ButeraMFM} $Q\sim 0.3$ and an average grain size
of 4 nm, which may be used as an estimation for the size of the pinning
inclusions. The wall width can be obtained\cite{Getzlaff} from $\delta =2%
\sqrt{A/K_{\perp }}\sim 16$ nm ($A\sim 10^{-6}$ erg/cm is the exchange
stiffness constant\cite{ButeraMFM}) giving $\beta _{0}\sim 0.14<1$ for the
studied films, which as an indication of weak pinning. We have then used $c=2
$ in Eq. \ref{Vac} and plotted the values of $V_{ac}$ as a function of film
thickness in Fig. \ref{FigVac}. We can observe that, within the experimental
error, there are no significant differences in the two approaches used for
calculating $V_{ac}$. Even more, the activation volume seems to be rather
constant for the different samples, with an average value $\overline{V_{ac}}%
=(1.37\pm 0.30)\times 10^{4}$ nm$^{3}$ which, for spherical volumes, is
equivalent to an average activation diameter $\overline{d_{ac}}=30\pm 3$ nm.
For the studied samples with $d>d_{cr}$ the activation diameter $\overline{%
d_{ac}}$ is larger than the grain size, which implies that although the
predominant interactions for $d>d_{cr}$ are dipolar-like, there seems to be
a positive intergranular exchange coupling which contributes to the
collective reversal of volumes larger than the grain size. This is
consistent with the fact that the interaction parameter $\alpha $ is
positive for $d=35$ nm, the first sample for which the stripe structure is
observed, and may also explain the positive ordinate in the $\alpha $ vs. $%
E_{S}$ curve of Fig. \ref{Fig6}. Note also that this value of the activation
diameter is of the same order than the film thickness or the stripe half
period when $d\gtrsim d_{cr}$, but is much shorter than the stripe length
(which is of the order of tens of micrometers), implying that the magnetic
volumes that reverse by thermal effects are considerably smaller than the
physical volume of the stripes. Our activation diameters are larger than
those reported in Refs. [\onlinecite{Jeong,Luo,Pernechele}] by a factor of
two or more (considering that the parameter $c$ appearing in Eq. \ref{Vac}
was taken as $c=1$ in those previously published papers). This is not
surprising due to the totally different microstructure between chemically
ordered and disordered samples and the larger exchange length in our
magnetically soft films.\cite{ButeraMFM}.

The activation volume obtained by the procedure described above may be
compared with the theoretical approach in the case of weak domain wall
pinning. In this case the activation energy to overcome the barrier depends
linearly\cite{Gaunt1,Gaunt2} on the magnetic field $H,$%
\begin{equation}
E_{a}=31\gamma \left( \delta /4\right) ^{2}\left( 1-H/H_{0}\right) ,
\end{equation}%
with $H_{0}$ the pinning field at zero temperature. Since the activation
volume is related to the field derivative of the activation energy, we can
write%
\begin{equation}
V_{ac}=\frac{dE_{a}}{dH}\frac{1}{cM_{s}}=\frac{31\gamma \left( \delta
/4\right) ^{2}}{H_{0}}\frac{1}{cM_{s}}\sim \frac{Q\delta ^{3}2\pi M_{s}}{%
H_{0}}.  \label{VacGaunt}
\end{equation}%
In the last formula we have used $c=2$ and $\gamma =K\delta .$ With this
equation it is possible to calculate the activation volume if the coercive
field at $T=0$ is known. We have discussed in Ref. [\onlinecite{Jonas}] that
at low temperatures there is an unexpected decrease in $H_{C}$ because
interface stress effects hinder the formation of stripes, so that a
reduction in $H_{C}$ occurs at low temperatures and a value for $H_{0}$ is
not experimentally accessible. However, we can still take the maximum value
of $H_{C}(T)=H_{C}^{\max }$ as a lower bound estimation for $H_{0}.$ Using
the data from Fig. \ref{FigHrem} and Eq. \ref{VacGaunt} we calculated $V_{ac}
$ for the set of samples with $d\geq 35$ nm and show the results in Fig. \ref%
{FigVac}. We can observe that the calculated values of $V_{ac}$ are
approximately independent of film thickness, with the exception of $d=35$
nm, a case that should be taken with extra care because a maximum in $%
H_{C}(T)$ was not observed in the studied temperature range. This form of
calculating $V_{ac}$ gave in all cases larger values than those obtained
using Eq. \ref{Vac}, approximately by a factor of two. The difference may be
due to the underestimation of $H_{0}$ or to an overestimation of the wall
width $\delta $. Apart from this relatively small discrepancy, the observed
experimental behavior is weakly dependent on film thickness, in accordance
with the prediction of Eq. \ref{VacGaunt}. \bigskip

Another experimental procedure for the estimation of the activation volume,
which does not need the explicit measurement of $\chi _{irr}$, is the
so-called "waiting time method".\cite{Collocott} This method is based on
time relaxation measurements of $M$ at different fields close to $H_{S}$,
the same curves that are used for the determination of $S(H).$ The model is
based on the assumption that both $S$ and $\chi _{irr}$ are relatively
constant for fields around $H_{S}.$ When $M(t,H)$ curves are plotted
together as a function of $\ln (t/t_{0})$ it can be shown that the following
relation is obeyed:%
\begin{equation}
\Delta H=H_{f}\,\ln (t_{i}).
\end{equation}%
If a horizontal line of constant $M$ is drawn, $\Delta H$ represents the
field distance between intersection points, and $t_{i}$ the time of
intersection. A plot of $\Delta H$ as a function of $\ln (t_{i})$ has a
slope $H_{f}$ from which $V_{ac}$ can be obtained using Eq. \ref{Hf}.

\begin{figure}[tbph]
\includegraphics[ width=9cm]{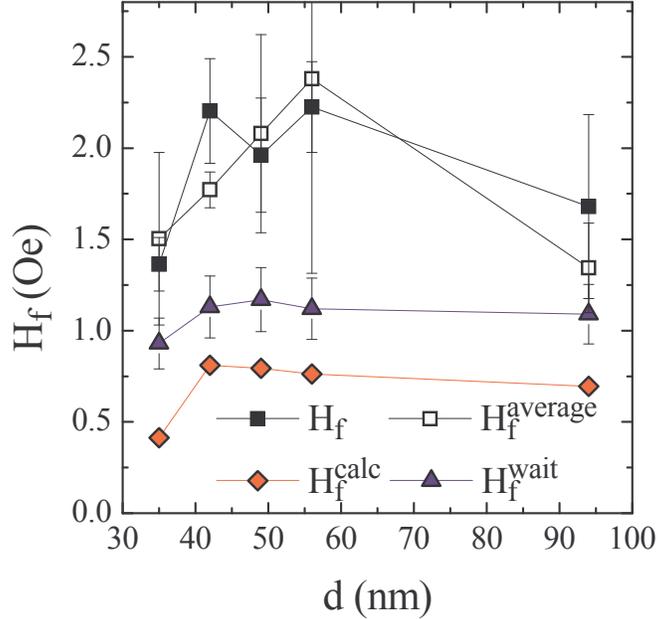}
\caption{Fluctuation field as a function of film thickness obtained from
Fig. \protect\ref{FigVac} and Eq. \protect\ref{Hf} (squares and diamonds)
and from the time relaxation of magnetization at different fields (waiting
time method, represented by triangles).}
\label{FigHf}
\end{figure}
In Fig. \ref{FigHf} we plotted the fluctuation fields obtained from the
previously calculated values of $V_{ac}$ and added the data deduced using
the waiting time method. Eventhough the error bars are relatively large, it
can be seen that these new values of $H_{f}$ are of the same order of
magnitude and relatively constant in the studied range of thicknesses,
consistent with those previously estimated using the remanence and viscosity
measurements. Following Ref. [\onlinecite{Wohlfarth}] we have tried to
correlate the values of $H_{f}$ with the coercivity $H_{C}$. According to
Wohlfarth there should be a power law relationship between both parameters, $%
H_{C}\varpropto H_{f}^{x}$, with $x$ in the range 0.5-1 depending on the
microstructure and the type of domain wall pinning of the system. Although
our data points fall close to those shown in Fig. 1 of Ref. [%
\onlinecite{Wohlfarth}] it was not possible to fit them using a power law
due to the reduced span of the coercivity and the fluctuation field values.

\bigskip

\section{Conclusions}

We have studied the role of magnetic interactions and thermally activated
processes in FePt alloy films as a function of film thickness. We have found
that when $d$ is larger than the critical thickness for the formation of a
structure of stripes with an antiparallel out of plane component of the
magnetization the interactions tend to be dipolar-like, while for $d<d_{cr}$
positive values of $\alpha $ are obtained$.$ This change is probably due to
the larger relative weight of the dipolar field present in the films with
stripe domains which arrange in a flux closure configuration that tends to
favor a demagnetized state. We have found that the large differences between
$H_{C}^{\mathrm{rem}}$ and $H_{\mathrm{IRM}}$ are mostly due to the
rotational anisotropy generated when the stripe structure is present. The
interaction parameter $\alpha $ becomes more negative with increasing
thickness which again is a consequence of the predominance of magnetostatic
demagnetizing effects for larger values of $d.$ We have shown that this
parameter is in close correlation with the surface demagnetizing energy,
confirming that dipolar interactions are predominant above the critical
thickness. Magnetic viscosity was also found to depend strongly on the
domain configuration. In thinner films relaxation seems to occur in discrete
steps while for $d>d_{cr}$ the usual logarithm behavior is found. $S(H)$ and
$\chi _{irr}(H)$ curves are a good estimation for the distribution of energy
barriers and also have a strong variation in the field width depending on
the domain structure. We finally estimated the values of the activation
volumes that reverse the magnetization assisted by thermal effects and found
that they are approximately independent of film thickness. The value of $%
\overline{d_{ac}}$ is almost an order of magnitude larger than the grain
size, evidencing that a relatively large number of grains is coupled by the
exchange interaction, but $\overline{d_{ac}}$ is considerably smaller than
the length of the stripes (which are several micrometers long), indicating
that the reversal occurs in small regions compared to the size of the
domains. Different methods of calculating the activation volumes and the
fluctuation fields yielded approximately the same results, supporting the
procedure used for the estimation of these parameters.

As far as we know, this is the first time that this kind of magnetic
measurements have been performed in chemically disordered FePt films in
which a transition in the domain structure occurs at a critical thickness.
We have clearly evidenced that strong changes in most variables accompany
the switch of the magnetic configuration from planar domains to parallel
stripes and gave an interpretation of the observed results. \bigskip

This was supported in part by Conicet under Grant PIP 112-200801-00245,
ANPCyT Grant PME \# 1070, and U.N. Cuyo Grant 06/C235, all from Argentina.
We would like to acknowledge very fruitful discussions with Dr. Emilio de
Biasi.

\end{document}